\documentclass{ws-procs975x65}

\usepackage{epsfig}
\newcommand{\Neel}{N\'{e}el }
\newcommand{\be}{\begin{equation}}
\newcommand{\ee}{\end{equation}}
\newcommand{\ben}{\begin{eqnarray}}
\newcommand{\een}{\end{eqnarray}}
\newcommand{\ra}{\rangle}
\newcommand{\la}{\langle}
\newcommand{\im}{{\rm i}}

\begin{document}

\title{FRUSTRATED QUANTUM ANTIFERROMAGNETS:\\ APPLICATION OF HIGH-ORDER COUPLED CLUSTER METHOD}

\author{J.~Richter$^*$, R.~Darradi$^*$, R.~Zinke$^*$}
\address{$^*$Institut f\"ur Theoretische Physik, Otto-von-Guericke
Universit\"at Magdeburg,\\
P.O.B. 4120, 39016 Magdeburg, Germany\\
www.uni-magdeburg.de/itp}

\author{R.F.~Bishop$^+$}
\address{$^+$School of Physics and Astronomy, The University of Manchester,\\
Sackville Street Building, P.O. Box 88, Manchester, M60 1QD, United Kingdom\\
www.physics.man.ac.uk}


\address{}

\begin{abstract}
We report on recent results  for strongly frustrated quantum $J_1$--$J_2$ antiferromagnets in 
dimensionality $d=1,2,3$ obtained by the coupled cluster method (CCM). 
We demonstrate that the CCM in high 
orders of approximation allows us to investigate quantum phase 
transitions driven 
by frustration and to discuss novel quantum ground states.     
In detail we consider the ground-state properties of (i) the 
Heisenberg spin-$1/2$ antiferromagnet on the cubic lattice in $d=1,2,3$, 
and use the results for the energy, the sublattice magnetization and the spin stiffness as a 
benchmark test for the precision of the method; 
(ii) coupled frustrated spin chains 
(the quasi-one-dimensional  $J_1$--$J_2$ model)
and discuss the influence of the quantum fluctuations and the interchain coupling on the incommensurate spiral 
state present in the classical model; 
(iii) the Shastry-Sutherland antiferromagnet on the square lattice; and
(iv) a stacked frustrated square-lattice Heisenberg antiferromagnet 
(the quasi-two-dimensional $J_1$--$J_2$ model), 
and discuss the influence of the interlayer coupling on the quantum paramagnetic ground-state phase 
that is present for the strictly two-dimensional model.
\end{abstract}


\bodymatter
\section{Introduction}\label{intro}
Strongly frustrated quantum magnets  
 have attracted much attention 
in recent years, both theoretically and experimentally.\cite{science,diep04,lnp04} 
In particular, quantum phase transitions  have very much become the 
focus of
interest (and see, e.g., Refs.~[\refcite{sachdev99,sachdev04,Richter04}]). 
At zero temperature, $T=0$, there are no thermal fluctuations present 
and the transitions between ground-state (GS) phases 
are driven purely by the interplay of quantum-mechanical fluctuations and 
competition between interactions (e.g., frustration).
In particular, novel quantum GS phases, such as valence-bond or 
spin-liquid phases  may appear\cite{Richter04,Misguich04} 
which do not possess semiclassical magnetic long-range order (LRO).
   
A basic model which shows strong quantum
fluctuations is the spin-$1/2$ pure Heisenberg
antiferromagnet (HAFM). 
The GS ordering of this quantum HAFM strongly depends on the
dimensionality.\cite{lnp04}
For example, unlike its three-dimensional (3d) and  two-dimensional (2d) model 
counterparts on the cubic or square lattice, respectively,
the one-dimensional (1d) pure HAFM (i.e., with nearest-neighbor antiferromagnetic 
bonds only, and all of equal strength)
does not have a N\'eel-ordered GS.\cite{Richter04,Mikeska04} 
Although the tendency to order is
more pronounced in 3d
quantum  spin systems than in lower-dimensional ones, a magnetically disordered
phase can also be
observed for 2d and 3d HAFM's when strong frustration is present, e.g., 
for the $J_1$--$J_2$ model on the square lattice (see, e.g., 
Refs.~[\refcite{chandra88,dagotto89,schulz,richter93,bishop98,
singh99,sushkov01,capriotti00,capriotti01,capriotti01a,siu01,singh03,sirker06}]
and references contained therein); 
or the 2d Shastry-Sutherland antiferromagnet;\cite{Shastry,Mila,
Miyahara,Lauchli,Weihong,Weihong02,Muller,Kawakami,Hajii,Subir,Miyahara03,
Richter04,rachid05} 
or, in three dimensions, for the HAFM on
the pyrochlore lattice \cite{canals98,moessner01} or on 
the stacked kagom\'e lattice.\cite{schmal04}

Besides the general interest in  frustrated  quantum antiferromagnets as 
rich examples of quantum many-body systems, 
there is a strong motivation for their theoretical study which is driven by  
the many recent experiments on 
quasi-low-dimensional materials that are well described by a frustrated  
spin-$1/2$ Heisenberg model. Among many others we mention here 
(i) the quasi-1d 
edge-sharing copper oxides 
like LiCuVO$_4$, LiCu$_2$VO$_2$  and NaCu$_2$O$_2$ 
\cite{ender,drechs1,drechs2,drechs3}, which were identified   
as frustrated (with ferromagnetic nearest-neighbor (NN) and antiferromagnetic 
next-nearest-neighbor (NNN) interactions) quantum helimagnets 
with an incommensurate spiral GS; 
(ii) the quasi-2d
{SrCu}$_2$({BO}$_3$)$_2$
\cite{Taniguchi,Kageyama99}, which is well described by the Shastry-Sutherland antiferromagnet, 
and which exhibits a gapped quantum paramagnetic GS; 
and (iii)
the quasi-2d  Li$_2$VOSiO$_4$
\cite{melzi00,rosner02} which can be described by the $J_1$--$J_2$ model on the square lattice.

The theoretical treatment of the frustrated 
quantum antiferromagnets is far from 
being trivial.   Though, surprisingly, one can find exact GS's  of a simple product 
nature in some exceptional cases,\cite{Shastry,maju,prl02} many of the 
standard many-body methods, 
such as quantum Monte Carlo techniques, may fail or become computationally infeasible 
to implement if frustration is present. 
Other methods, such as density-matrix renormalization group (DMRG) or exact 
diagonalization techniques, are essentially 
restricted to low-dimensional systems, at least for the present.      
Hence, there is considerable interest in any method that can deal with frustrated 
spin systems in any number of dimensions, including magnetic systems with incommensurate
spiral GS's. 
A method  fulfilling this requirement is the coupled cluster method (CCM).
This approach, introduced many years ago by Coester and K\"ummel,\cite{coest}
is one of the most universal and most 
powerful methods of quantum many-body theory (and for a review of which  
see, e.g., Ref.~[\refcite{bish_rev}]).
The CCM has previously been 
applied to various  quantum spin systems with much
success.\cite{rog_her90,bishop91,bishop94,bursill,bishop98,zeng98,
krueger00,bishop00,krueger01,ivanov02,
rachid04,Farnell04,rachid05,Farnell05,rachid06,schmal06}
The application to frustrated  spin systems was 
started in the 1990's,\cite{bursill,bishop98} and has been developed in 
more recent years 
to the point where it has become a powerful tool in this 
field by including higher orders of approximations in a well-defined 
truncation scheme.\cite{krueger00,krueger01,
ivanov02,rachid05,Farnell05,schmal06}

In this paper, we review some recent applications of the CCM
to the generic frustrated quantum spin-$1/2$ Heisenberg model 
\begin{equation} \label{HM}
H = \sum_{ i,j } J_{ij}\; {\bf s}_i \cdot {\bf s}_j \; ,
\end{equation}
where the ${\bf s}_i$ are the spin operators fulfilling ${\bf s}_i^2=3/4$, and the $J_{ij}$
are the (competing) exchange coupling parameters.

The paper is organized as follows.  In Sec.~\ref{ccm}
we illustrate the main features of the CCM, paying special attention to the methodological  
aspects relevant for the application of the CCM to frustrated magnets.
In Sec.~\ref{unfrust} we report (as a benchmark test)
high-order CCM results for standard unfrustrated lattices in 
$d=1,2,3$ and compare them with other accurate methods. In Sec.~\ref{1d} we discuss 
the influence of quantum fluctuations on GS spiral ordering for quasi-1d $J_1$--$J_2$ 
spin systems, and in Sec.~\ref{shast} we report on a CCM treatment of the 2d 
Shastry-Sutherland model. Finally, Sec.~\ref{3d} contains 
a discussion of the GS ordering of a 
stacked frustrated square-lattice HAFM 
(the quasi-2d  $J_1$--$J_2$ model).

\section{The Coupled Cluster Method}\label{ccm}
The CCM formalism is now briefly outlined. For further details the
interested reader is referred to Refs.~[\refcite{bishop91,bishop00,Farnell04,zeng98,krueger00,Farnell05,rachid06}].
The starting point for the CCM calculation is the choice of a normalized reference or model state
$|\Phi\rangle$, together with a set of (mutually commuting) multi-configurational creation operators $\{ C_L^+ \}$ and 
the corresponding set of their Hermitian adjoints $\{ C_L \}$,
\begin{equation}
\label{eq2.1} \langle \Phi|C_L^+ = 0 = C_L|\Phi \rangle 
\quad \forall L\neq 0 \; , \quad C_0^+\equiv 1 \; ; \quad[C_L^+,C_J^+] = 0 =[C_L,C_J] \; .
\end{equation}
The operators $C_L^+$ ($C_L$) are defined over a complete set of many-body configurations 
denoted by the set of set-indices $\{ L \}$. 
With the set $\{|\Phi\rangle, C_L^+\}$ the CCM parametrizations of the exact ket and bra GS eigenvectors
$|\Psi\ra$ and $\la\tilde{\Psi}|$ of our many-body system are then given  by
\begin{equation}\label{eq5} 
|\Psi\ra=e^S|\Phi\ra \; , \mbox{ } S=\sum_{L\neq 0}a_LC_L^+ \; ; \quad
\la \tilde{ \Psi}|=\la \Phi |\tilde{S}e^{-S} \; , \mbox{ } \tilde{S}=1+ \sum_{L\neq 0}\tilde{a}_LC_L \; .
\end{equation}
The CCM correlation operators, $S$ and $\tilde{S}$, contain the correlation coefficients, $a_L$ and $\tilde{a}_L$, 
which have to be calculated.  Once known, all GS properties of the many-body system can clearly be found 
in terms of them.  
To find the GS correlation coefficients $a_L$ and $\tilde{a}_L$,
we simply require that the GS energy expectation value $\bar H=\la\tilde\Psi|H|\Psi\ra$ is a minimum with respect
to the entire set $\{ a_L , \tilde{a}_L \}$, which leads to the GS CCM ket-state and bra-state
equations
\begin{equation}
\label{eq6}
\langle\Phi|C_L^-e^{-S}He^S|\Phi\rangle = 0 \; ,\mbox{ }\langle\Phi|{\tilde S}e^{-S}[H, C_L^+]e^S|\Phi\rangle = 0 \quad \forall L\neq 0 \; .
\end{equation}

For frustrated spin systems  an appropriate choice for the CCM model state $|\Phi\rangle$ is often  a 
classical spiral spin state, 
(i.e., pictorially, $|\Phi\rangle =|\uparrow\nearrow\rightarrow\searrow\downarrow\swarrow\cdots\rangle$), 
which is characterized 
by a pitch angle $\alpha$.  Such states include the \Neel state, for which $\alpha=\pi$. 
In the quantum model the pitch angle may be modified by quantum fluctuations. Hence, we do not choose the 
classical  result for the pitch angle but, rather, we consider it as a free parameter in the CCM 
calculation, which is to be determined by minimization of the CCM GS energy.

In order to find an appropriate set of creation operators it is convenient to perform 
a rotation of the local axes of each of the spins, such that all spins in the reference state
align in the negative $z$-direction. 
This rotation by an appropriate angle $\delta_i$ of the spin on lattice site $i$ 
is equivalent to the spin-operator transformation
\begin{equation}
\label{eq3} s_i^x = \cos\delta_i {\hat s}_i^x+\sin\delta_i {\hat s}_i^z \; ; \quad s_i^y = {\hat s}_i^y\;;\quad
s_i^z = -\sin\delta_i {\hat s}_i^x+\cos\delta_i {\hat s}_i^z \; .
\end{equation}
In this new set of local spin coordinates 
the reference state and the corresponding creation operators $C_L^+$ are given by
\begin{equation}
\label{set1} |{\hat \Phi}\ra = |\downarrow\downarrow\downarrow\downarrow\cdots\rangle \; ; \quad C_L^+ 
= {\hat s}_i^+ \, , \, {\hat s}_i^+{\hat s}_j^+ \, , \, {\hat s}_i^+{\hat s}_j^+{\hat s}_k^+ \, ,\, \ldots \; ,
\end{equation}
where the indices $i,j,k,...$ denote arbitrary lattice sites.
In the new coordinates the initial Heisenberg Hamiltonian of Eq.~(\ref{HM}) becomes  
\begin{eqnarray}\label{produkt_trafo}
\nonumber {\hat H}= &&\frac{1}{2}\sum_{i,j} J_{ij} \left\{ [\cos(\alpha_{ij})+1]({\hat s}_i^+{\hat s}_{j}^-+{\hat s}_i^-{\hat s}_{j}^+) +
[\cos(\alpha_{ij})-1]({\hat s}_i^+{\hat s}_{j}^++{\hat s}_i^-{\hat s}_{j}^-)\right. \\
&&+\left. 2\sin(\alpha_{ij})[{\hat s}_i^+{\hat s}_{j}^z-{\hat s}_i^z
{\hat s}_{j}^++{\hat s}_i^-{\hat s}_{j}^z-{\hat s}_i^z{\hat s}_{j}^-] +4\cos(\alpha_{ij}){\hat s}_i^z{\hat s}_{j}^z \right \},
\end{eqnarray}
where $\alpha_{ij}\equiv\delta_{j}-\delta_i$ is the angle between the two interacting spins, and
${\hat s}_i^{\pm}\equiv {\hat s}_i^x\pm \im {\hat s}_i^y$.
In the case of the N\'eel model state $\hat H$ becomes much simpler, so that, for example, for the case of NN bonds only, we have
${\hat H} = -J\sum_{\langle i,j\rangle} \left({\hat s}_i^+{\hat s}_{j}^+ +{\hat s}_i^-{\hat s}_{j}^- +2{\hat s}_i^z{\hat s}_{j}^z\right)$.

For the ensuing discussion of the GS properties we concentrate on the GS energy $E$,
the order parameter $m$ and the spin stiffness ${\rho_s}$.
Within the CCM scheme we have 
$E=\la\Phi|e^{-S}He^S|\Phi\ra$ and 
$m= -\frac{1}{N}\la\tilde{\Psi}|\sum_{i=1}^N{\hat s}_i^z|\Psi\ra$.
The spin stiffness ${\rho_s}$ can be calculated 
by imposing a twist on the order parameter of a magnetically 
long-range ordered system along a given direction, i.e., 
\begin{equation}
\label{stiff1}\frac{E({\theta})}{N}=\frac{E({\theta=0})}{N}+\frac{1}{2}{\rho_s}{\theta^2}+{\cal O}({\theta^4}) \; ,
\end{equation}
where $E({\theta})$ is the GS energy as a function of the twist angle $\theta$, 
and $N$ is the number of sites (and where the interested reader is referred to 
Ref.~[\refcite{rachid06}] for further details). 

The CCM formalism is exact if we take into account all possible multispin configurations in 
the correlation operators $S$ and $\tilde S$.  However, in general, this is  
impossible to do in practice for  a quantum many-body system. 
Hence, it is necessary to use approximation 
schemes in order to truncate the expansions of $S$ and $\tilde S$ in Eq.~(\ref{eq5}) in any practical calculation.  
A quite general  approximation scheme is the so-called SUB$n$-$m$ approximation. In this approximation all correlations 
in the correlation operators $S$ and $\tilde S$ are taken into account which span a range of no more than 
$m$ contiguous sites and contain only  $n$ or fewer spins. 
In most cases  the SUB$n$-$n$ scheme is used (i.e., with $n=m$), and it is then (for spin-$1/2$ systems) called the LSUB$n$ scheme. 
To find all different fundamental configurations entering $S$ and $\tilde S$ at a given level of LSUB$n$ approximation 
we use lattice symmetries and, where possible, any exact 
conservation laws. 

Since the LSUB$n$  approximation becomes exact in the limit 
$n \to \infty$, it is useful to extrapolate the 'raw' LSUB$n$  results to the limit $n \to \infty$.  
Although an exact scaling theory for the LSUB$n$ results is not known, there is some empirical experience 
\cite{bishop00,zeng98,krueger00,rachid05,schmal06} indicating how the physical quantities for spin models 
might scale with $n$. 
For the GS energy we employ\cite{bishop00,krueger00}
\be \label{scal_e}
E(n) = a_0 + a_1\frac{1}{n^2} + a_2\left(\frac{1}{n^2}\right)^2 \, .
\ee
Furthermore, we note that it may be useful to discard the LSUB2 results for the extrapolation, 
because generally they fit poorly to the asymptotic behavior\cite{bishop00}.  
For the order parameter  and the stiffness 
one  utilizes \cite{bishop00,rachid06} 
an extrapolation law with leading power $1/n$, i.e., 
\be \label{scal_m1}
A(n) = b_0 + b_1\frac{1}{n} + b_2\left(\frac{1}{n}\right)^2 \, .
\ee
However, there is some experience that when applied  
to systems showing an order-disorder quantum phase transition 
this kind of extrapolation tends to overestimate the parameter region where magnetic LRO
exists, i.e.  
to yield too large critical values for the 
exchange parameter driving the transition.\cite{krueger00,rachid04,rachid05,rachid06} 
The reason for such behavior might         
derive from the change of the scaling near a critical point.
Hence, in addition to the extrapolation rule of Eq.~(\ref{scal_m1}) for the order parameter $m$, we also 
use a leading 'power-law' extrapolation\cite{bishop00,rachid05,rachid06} given by
\be \label{scal_m2}
 m(n)=c_0+c_1\left(\frac{1}{n}\right)^{c_2} \, ,
\ee
where the leading exponent $c_2$ is determined directly from 
the LSUB$n$ data.

\section{The CCM for the Pure HAFM on Cubic Lattices for $d=1,2,3$}\label{unfrust}

During the last few years the running time and memory requirements of the original CCM code 
have been considerably improved.\cite{Farnell05,cccm} Consequently, it is now  possible to run higher 
levels of approximation by using an improved parallelization procedure.
In this Section we present a collection of CCM results for the (unfrustrated) pure HAFM 
(i.e., with the Hamiltonian of Eq.~(\ref{HM}) with $J_{ij} = 1$ for NN bonds, and $J_{ij} = 0$ otherwise) 
on some basic lattices, and compare them  with the most accurate 
results obtained by other methods.  While some of the CCM results have already been 
published elsewhere,\cite{bishop00,rachid06} we also present here the new 
unpublished results from higher levels of approximation. The results are shown in Tables \ref{tabelle1}, \ref{tabelle2} 
and \ref{tabelle3}, in which the entries shown in boldface are the new ones.

\begin{table}
\tbl{Data for the spin-$1/2$ linear-chain pure HAFM.
$N_f$ is the number of fundamental configurations for the N\'eel reference  state, 
$E/N$ is the GS energy per spin and $m$ is the sublattice magnetization. 
The LSUB$n$ results are extrapolated using Eq.~(\ref{scal_e}) for $E/N$ and  
Eq.~(\ref{scal_m1}) (in parentheses, Eq.~(\ref{scal_m2}))for $m$.}
{\begin{tabular}{@{}|c|c|c|c|@{}}\hline 
{\bf linear chain}        &$N_f$          &$E/N$          &$m$            \\ \hline
LSUB2                     &1              &-0.41667       &0.33333       \\\hline
LSUB4                     &3              &-0.43627       &0.24839        \\\hline
LSUB6                     &9              &-0.44002       &0.20789        \\\hline
LSUB8                     &26             &-0.44137       &0.18297        \\\hline
LSUB10                    &81             &-0.44200       &0.16562        \\\hline
LSUB12                    &267            &-0.44234       &0.15263        \\\hline
{\bf LSUB14}              &{\bf 931}      &{\bf -0.44255} &{\bf 0.14240}  \\\hline
{\bf LSUB16}              &{\bf 3362}     &{\bf -0.44269} &{\bf 0.13408}  \\\hline
{\bf Extrapolated CCM}    &-              &{\bf -0.44315} &{\bf 0.07737} ({\bf -0.01086}) \\\hline
Bethe ansatz\cite{bethe}  &-              &-0.44315       &0.0            \\\hline
\end{tabular}}
\label{tabelle1}
\end{table}

\begin{table}
\tbl{Data for the spin-$1/2$ square-lattice pure HAFM.
$N_f$ is the number of fundamental configurations for the N\'eel (in parenthesess, for the spiral) reference  state, 
$E/N$ is the GS energy per spin, $m$ is the sublattice magnetization and $\rho_s$ is the  spin stiffness.
The LSUB$n$ results are extrapolated using Eq.~(\ref{scal_e}) for $E/N$, 
Eq.~(\ref{scal_m1}), (in parentheses, Eq.~(\ref{scal_m2})) for $m$ and Eq.~(\ref{scal_m1}) for $\rho_s$.}
{\begin{tabular}{@{}|c|c|c|c|c|@{}}\hline 
{\bf square lattice}      &$N_f$                 &$E/N$          &$m$                            &$\rho_s$    \\ \hline
LSUB2                     &1 (3)                 &-0.64833       &0.42071                        & 0.2574 \\ \hline
LSUB4                     &7 (40)                &-0.66366       &0.38240                        & 0.2310 \\ \hline
LSUB6                     &75 (828)              &-0.66700       &0.36364                        & 0.2176 \\ \hline
LSUB8                     &1287 (21124)          &-0.66817       &0.35242                        & 0.2097 \\ \hline
{\bf LSUB10}              &{\bf 29605 (586787)}  &{\bf -0.66870} &{\bf 0.34483}                  & -           \\ \hline
{\bf Extrapolated CCM}    &-                     &{\bf -0.66936} &{\bf 0.31024} ({\bf 0.28073 }) & 0.1812  \\ \hline
3rd order SWT\cite{swts}  &-                     &-0.66931       &0.3069                         &0.1747        \\ \hline
QMC\cite{qmc}             &-                     &-0.66944       &0.3070                         &0.1852        \\ \hline
\end{tabular}}
\label{tabelle2}
\end{table}

\begin{table}
\tbl{Data for the spin-$1/2$ simple-cubic lattice pure HAFM.
$N_f$ is the number of fundamental configurations for the N\'eel (in parenthesess, for the spiral) reference  state, 
$E/N$ is the GS energy per spin, $m$ is the sublattice magnetization and $\rho_s$ is the  spin stiffness.
The LSUB$n$ results are extrapolated using Eq.~(\ref{scal_e}) for $E/N$, 
Eq.~(\ref{scal_m1}), (in parentheses, Eq.~(\ref{scal_m2})) for $m$ and Eq.~(\ref{scal_m1}) for $\rho_s$.}
{\begin{tabular}{@{}|c|c|c|c|c|@{}}\hline 
{\bf cubic lattice}       &$N_f$                 &$E/N$          &$m$                            &$\rho_s$       \\ \hline
LSUB2                     &1 (4)                 &-0.89076       &0.45024                        & 0.2527   \\ \hline
LSUB4                     &9 (106)               &-0.90043       &0.43392                        & 0.2416  \\ \hline
LSUB6                     &181 (5706)            &-0.90180       &0.42860                        & 0.2380   \\ \hline
LSUB8                     &{\bf 8809 (444095)}   &{\bf -0.90214} &{\bf 0.42626}                  &-              \\ \hline
{\bf Extrapolated CCM}    &-                     &{\bf -0.90247} &{\bf 0.42054} ({\bf 0.42141})  & 0.2312         \\ \hline
3rd order SWT\cite{swtcneu}     &-                     &-0.9025         &0.4227                         & 0.2343        \\ \hline
\end{tabular}}
\label{tabelle3}
\end{table}

\section{The Quasi-One-Dimensional $J_1$--$J_2$ Model}
\label{1d}
One-dimensional quantum spin systems like the 
frustrated $J_1$--$J_2$-model have attracted much 
attention over many years.\cite{Mikeska04} 
The physics of such 
quantum spin systems is often remarkably different from that of their corresponding 
classical counterparts, with a rich variety of different quantum GS's. 
As already mentioned in Sec.~\ref{intro}, novel edge-sharing copper oxides 
like LiCuVO$_4$ and NaCu$_2$O$_2$,\cite{ender,drechs1,drechs2,drechs3} that were identified   
as quantum helimagnets with ferromagnetic NN  and frustrating antiferromagnetic NNN 
bonds, have stimulated a great deal of renewed interest in frustrated 
Heisenberg chains.  In these materials a finite interchain coupling is also present, which in turn 
may lead to a low-temperature phase transition to a magnetically ordered phase.

\begin{figure}[t]
\begin{center}
\psfig{file=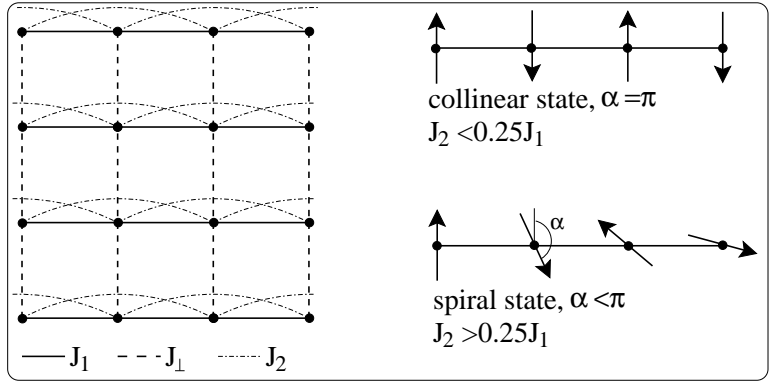,width=4.0in}
\end{center}
\caption{Illustration of the quasi-1d $J_1$--$J_2$ model and the reference states $|\Phi\rangle$ 
used for our CCM calculations.}
\label{themodel}
\end{figure}

Therefore, we consider in this section the following quasi-1d spin-$1/2$ $J_1$--$J_2$ Heisenberg model, 
\begin{equation}
\label{eq1.1}
H=\sum_n\Bigg(J_1\sum_{i}{\bf s}_{i,n} \cdot {\bf s}_{i+1,n}
+J_2 \sum_{i}{\bf s}_{i,n} \cdot {\bf s}_{i+2,n}\Bigg)
 + J_\perp \sum_{i,n} {\bf s}_{i,n} \cdot {\bf s}_{i,n+1} \; ,
\end{equation}
where $n$ labels the chains,
$J_\perp$ is the interchain coupling, $J_1$ is the in-chain 
NN coupling  
and $J_2$ the in-chain NNN coupling,
as shown in Fig.~\ref {themodel}.
We fix $J_1$ to $J_1=1$ (for the antiferromagnetic case) or to $J_1=-1$ 
(for the ferromagnetic case) and consider $J_2\geq 0$ (which is frustrating in 
both cases) and $J_{\perp} \geq 0$.  The case $J_{\perp}= 0$ corresponds to the strictly 1d case.

The classical system is characterized by a second-order
transition from a  collinear phase to a non-collinear spiral phase at $|\frac{J_2}{J_1}|=0.25$. 
For $J_2\ge |J_1|$ the  classical spiral (pitch) angle $\alpha_{\rm cl}$ is given by 
$\alpha_{\rm cl}=\arccos\left( -0.25 J_1/J_2\right)$.
Note that in the classical model neither the pitch angle $\alpha_{\rm cl}$ nor the transition point 
$|\frac{J_2}{J_1}|=0.25$ depends on the interchain coupling $J_{\perp}$.

From the experimental point of view it is of interest to discuss the
influence of quantum fluctuations on the pitch angle.  Furthermore the question arises 
whether in the quantum model the interlayer coupling $J_\perp$ 
does or does not influence either or both of the pitch angle and the transition point.
While the influence of quantum fluctuations on the pitch angle for $J_\perp=0$ has been discussed previously 
in the literature,\cite{bursill,white96,aligia00,dima06} 
this question has not been considered for finite $J_\perp > 0$ so far.
Furthermore,  any effect which a finite $J_\perp$ may have  
on the transition point has also not been considered in the literature. 
To discuss this point we use 
the CCM for the  model of Eq.~(\ref{eq1.1}) at the same level of 
approximation (viz., SUB2-3) 
as in Ref.~[\refcite{bursill}]. 
In that paper it was demonstrated that the SUB2-3 approximation for strictly 1d systems 
leads to results of comparable 
accuracy to those obtained using the DMRG method.  For the SUB2-3 approximation 
the relevant CCM equations can be found  in closed analytical form, even for 
nonzero values of $J_\perp$.

\def\figsubcap#1{\par\noindent\centering\footnotesize(#1)}

\begin{figure}[ht]
\begin{center}
  \parbox{2.4in}{\epsfig{figure=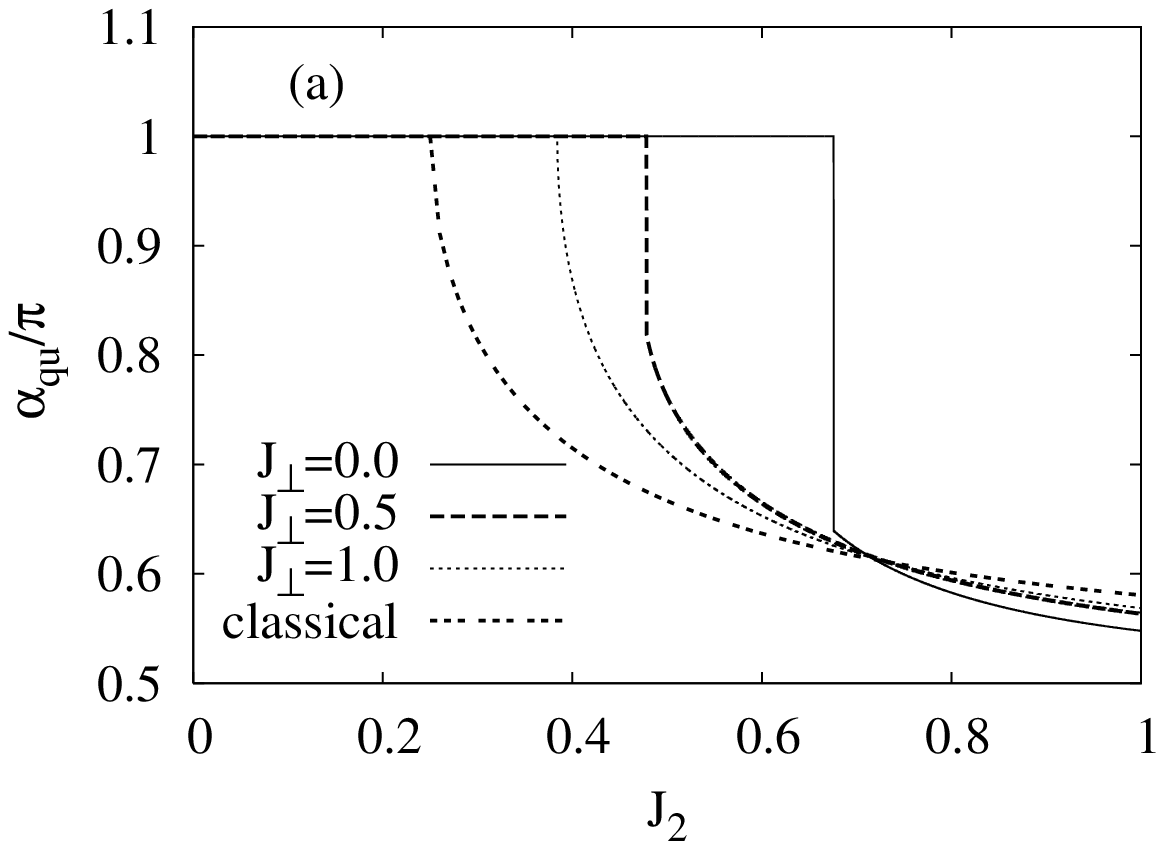,width=2.6in}}
  \hspace*{4pt}
  \parbox{2.4in}{\epsfig{figure=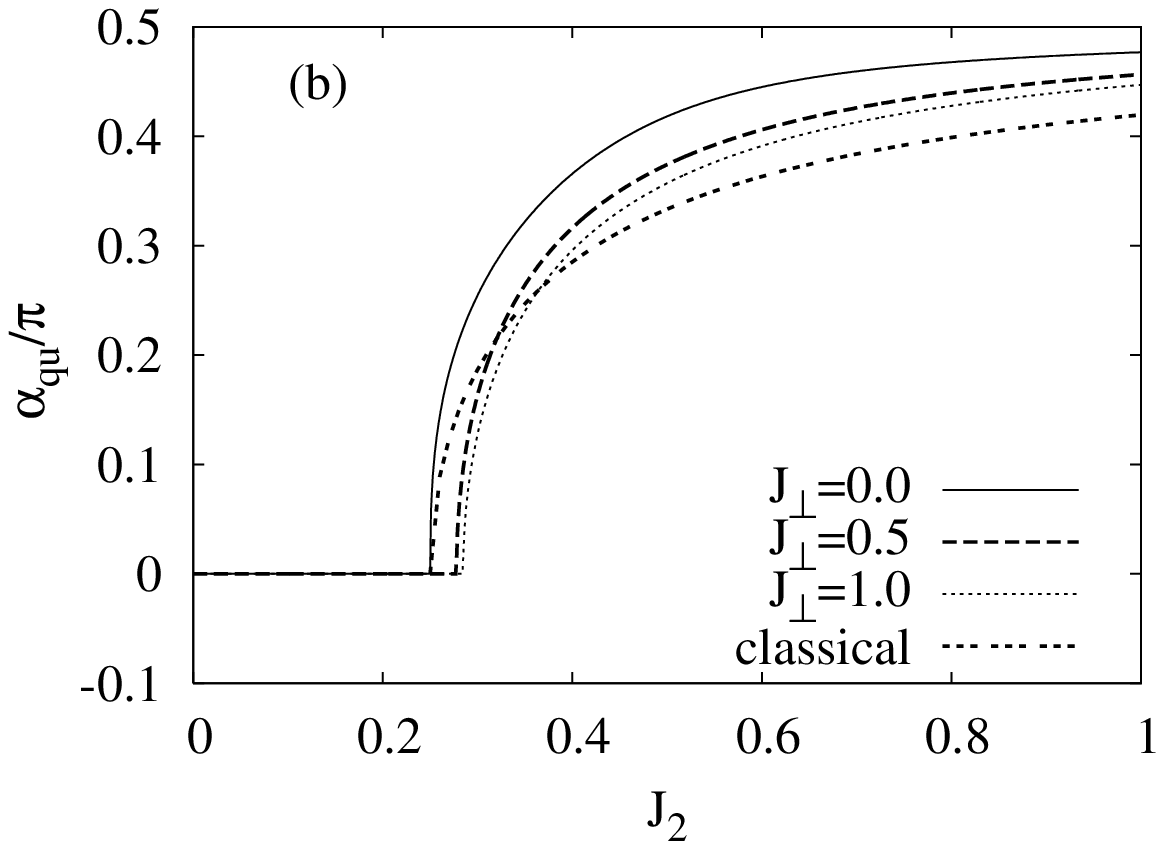,width=2.6in}}
\end{center}
\caption{The quantum pitch angle $\alpha_{\rm qu}$ versus $J_2$  for the two 
cases (a) $J_1=1$ and (b) $J_1=-1$ in the CCM SUB2-3 approximation, compared 
with the classical result.}
  \label{fig3a}
\end{figure}

\def\figsubcap#1{\par\noindent\centering\footnotesize(#1)}
\begin{figure}[ht]
\begin{center}
  \parbox{2.4in}{\epsfig{figure=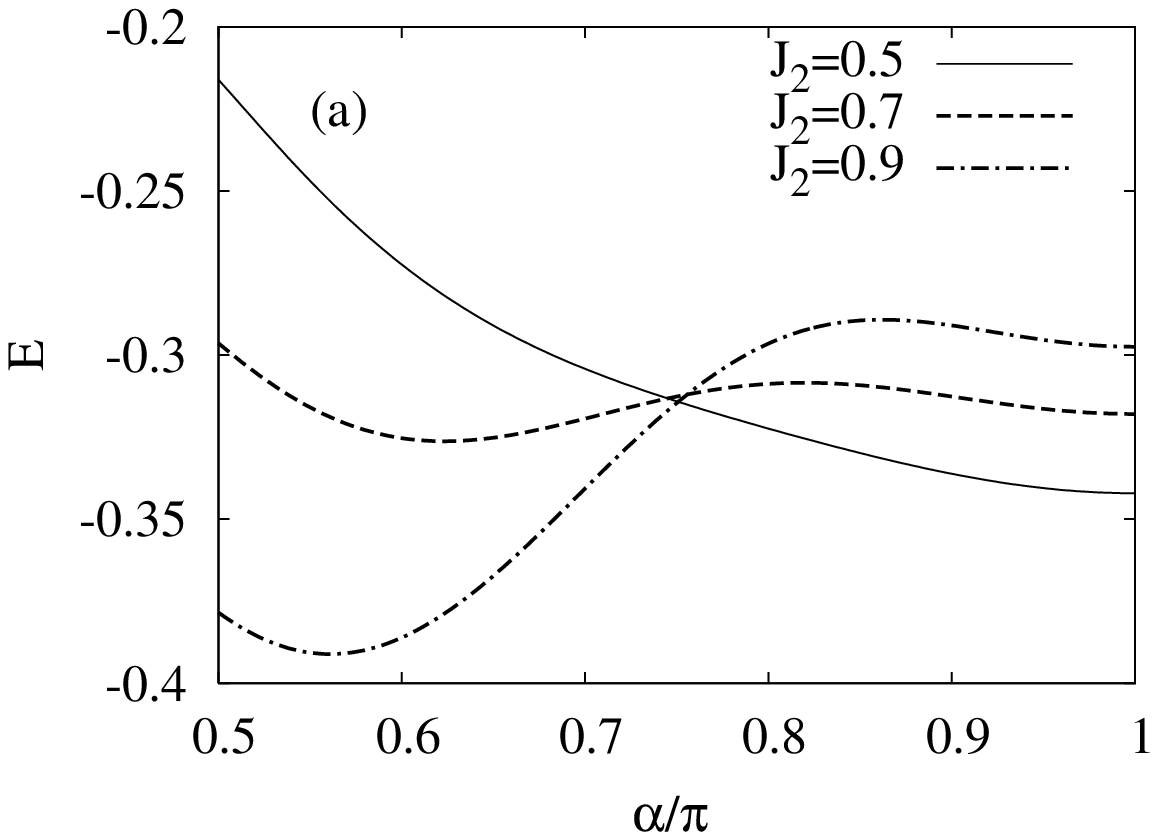,width=2.6in}}
  \hspace*{4pt}
  \parbox{2.4in}{\epsfig{figure=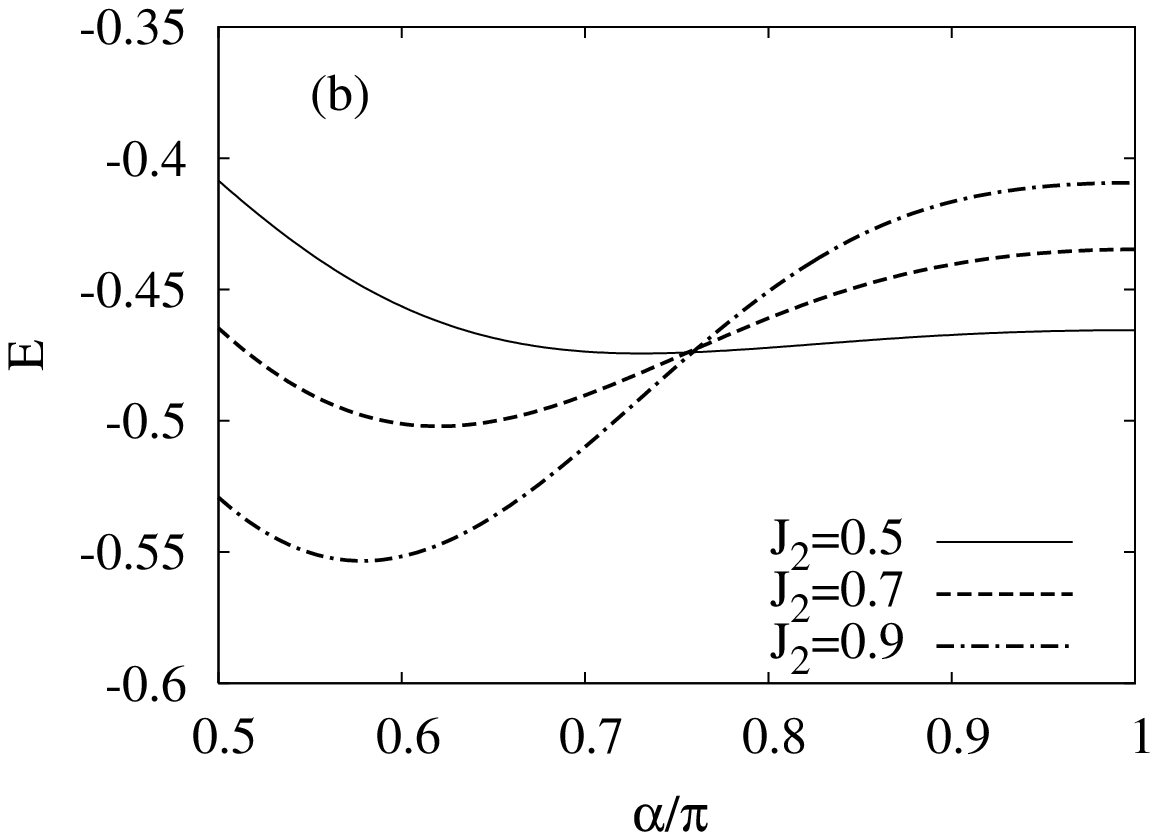,width=2.6in}}
\caption{Ground-state energy per spin in the CCM SUB2-3 approximation as a function
of the pitch angle $\alpha$ for different (fixed) values of the NNN 
frustration strength $J_2$ and antiferromagnetic NN bonds with $J_1=1$, for the 
two cases (a) $J_{\perp}=0$ and (b) $J_{\perp}=0.7J_1$. 
The quantum pitch angle $\alpha_{\rm qu}$ shown in Fig.~\ref{fig3a} corresponds to the absolute minimum 
of the energy $E$.}
  \label{fig1a}
\end{center}
\end{figure}

As quantum fluctuations may lead to a `quantum'
pitch angle $\alpha_{\rm qu}$  that is different from the classical value $\alpha_{\rm cl}$, we consider 
the pitch angle in the reference state as a free parameter. 
We then determine $\alpha_{\rm qu}$ 
by minimizing the CCM energy $E(\alpha)$ with respect to $\alpha$. 
Some results for the pitch angle $\alpha_{\rm qu}$ of the quantum model are 
presented  in Fig.~\ref{fig3a}. We have checked explicitly, that  for 
$J_\perp = 0$ our results coincide with those of Ref.~[\refcite{bursill}].  
(We note that due to the different definitions of the Hamiltonian 
parameters that agreement is not obvious).     
For  antiferromagnetic $J_1$ and weak interchain coupling $J_\perp$ 
we find, contrary to the behavior of the classical system, a first-order transition  
from a collinear phase to a spiral phase, i.e., the quantum pitch angle  jumps from $\alpha_{\rm qu}=\pi$ 
(in the collinear phase)
to $\alpha_{\rm qu}< \pi$ (in the spiral phase). This is illustrated in Fig.~\ref{fig1a} by the energy 
curves $E(\alpha)$, which show a typical first-order scenario for 
$J_\perp=0$ compared with a typical second-order scenario 
for $J_\perp=0.7J_1$. Furthermore, the transition point is both shifted to larger 
values of $J_2 > 0$ than in the classical case 
(and cf. Refs.~[\refcite{bursill,white96,aligia00}]), and   
does now depend on $J_{\perp}$. 
For increasing $J_{\perp}$ the transition point is shifted towards the classical value (but even for $J_\perp=J_1$ it remains 
significantly above the classical value), and for large enough $J_\perp$ the transition  between  the collinear 
and the spiral phase  
becomes continuous. 

For  ferromagnetic $J_1 = -1$ one has, in complete analogy to the classical system, a second-order transition  
from the collinear to a spiral phase  at $J_2=0.25$ when $J_\perp=0$, which is in agreement with 
earlier considerations\cite{dima06,krueger01}.
However, for $J_2 > 0.25 |J_1|$ the quantum pitch angle deviates from the classical value and 
depends on the interlayer coupling $J_\perp$.
There is also  a shift of the transition point for increasing values of 
$J_{\perp}$ towards larger values of $J_2$. 

A common feature for both cases (i.e., $J_1=-1$ and $J_1=+1$) is, that for increasing 
values of $J_2$ the quantum pitch angle  approaches 
its limiting value 
$\pi/2$ much faster than for the classical model.

The general behavior discussed above can be qualitatively related to the strength of quantum fluctuations 
(and see, e.g., the discussion in Refs.~[\refcite{krueger00,krueger01,rachid05})], which 
themselves depend on the 
values of the exchange parameters of the model.   
In general, quantum fluctuations tend to stabilize collinear phases. Consequently, for the model with the strongest
fluctuations, namely the strictly 1d antiferromagnetic model ($J_\perp=0, J_1=1$), the transition to the non-collinear GS 
takes place for largest $J_2$ (cf. Fig.~\ref{fig3a}a).    
Increasing the interchain coupling then reduces the strength  of the quantum fluctuations, which leads to a shift of the transition 
point to smaller values of $J_2$.
In the case of the strictly 1d chain with ferromagnetic $J_1=-1$ and $J_2 < 0.25|J_1|$, the GS is the fully polarized ferromagnetic state, which 
does not exhibit any  quantum fluctuations. As a result the transition point is the same as in the classical model, but the quantum pitch angle 
starts to deviate from the classical value immediately if $J_2 > 0.25|J_1|$.

Finally, we emphasize that we did not speculate on possible spiral LRO. For strictly 1d systems it is clear 
that there is no magnetic LRO and 
the quantum pitch angle discussed above corresponds to incommensurate short-range correlations.

\section{A Frustrated Heisenberg Antiferromagnet on the Square Lattice: The Shastry-Sutherland Model}
\label{shast}
\begin{figure}[htb]
\begin{center}
\epsfig{file=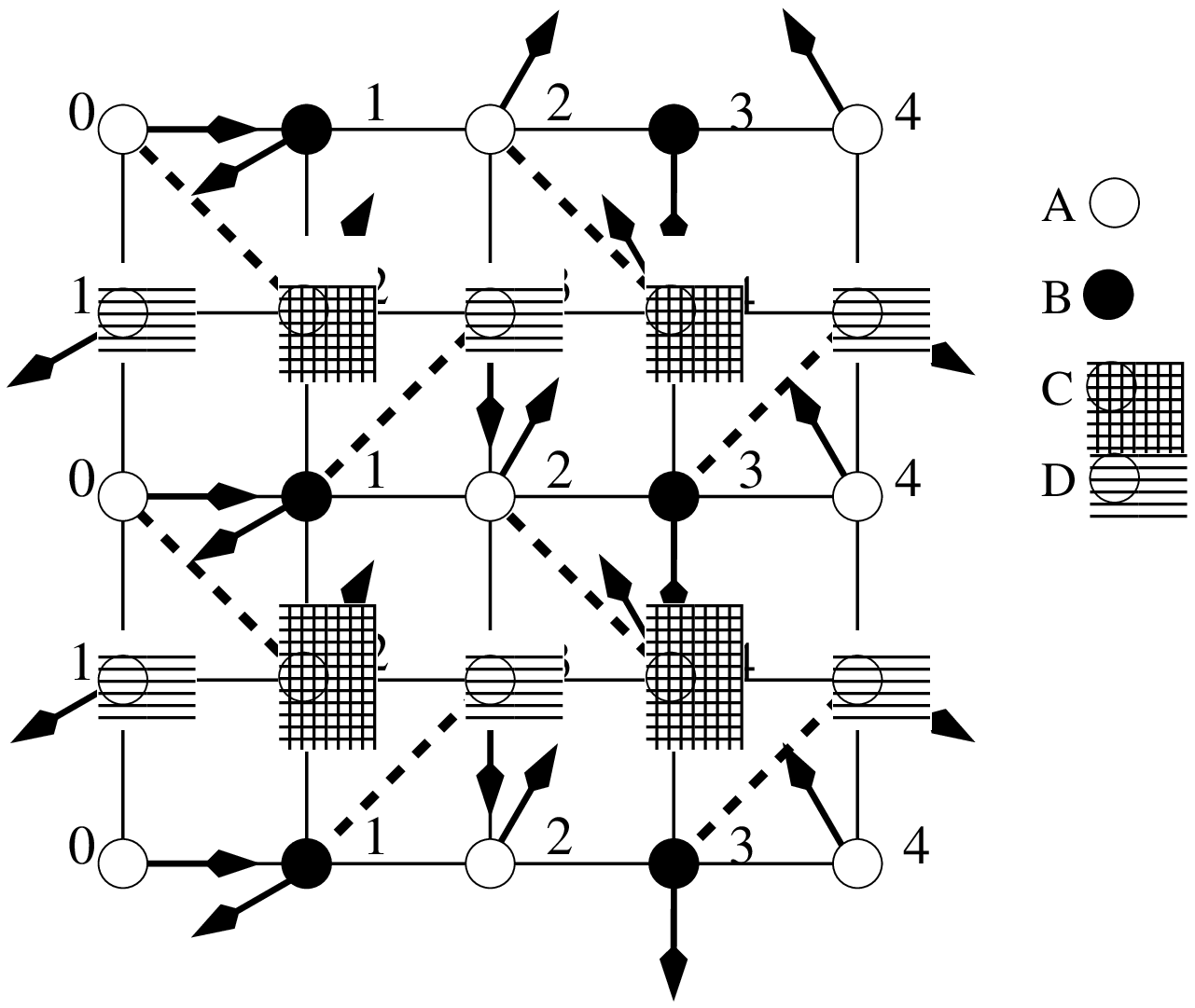,scale=0.4,angle=0}
\end{center}
\caption{\label{fig1} Illustration of the 
Shastry-Sutherland model with antiferromagnetic NN bonds $J_1=1$ (solid lines) and NNN bonds
$J_2$ (dashed lines), together with its classical spiral state.
The spin orientations at lattice sites $n$ are given by the angles
$\theta=n\alpha_{\rm cl}$, where $n=0, 1, 2,\dots,$ and $\alpha_{\rm cl}$
is the characteristic pitch angle of the classical spiral state. 
The state is shown for $\alpha_{\rm cl}=5\pi/6$ and
$n=0, 1, \dots ,5$.}
\end{figure}

As already mentioned in Sec.~\ref{intro}, 
the Shastry-Sutherland model, first 
introduced some 25 years ago,\cite{Shastry} has attracted much attention in connection with experiments on 
{SrCu}$_2$({BO}$_3$)$_2$.   
The model is characterized   
by a special arrangement of frustrating NNN
$J_2$ bonds  on the square lattice, as shown in Fig.~\ref{fig1}.
Its  Hamiltonian reads 
\begin{equation}
\label{eq1}
H = J_1\sum_{\langle i,j \rangle}{\bf s}_i \cdot {\bf s}_j 
+ J_2\sum_{\{i, k\}}{\bf s}_i \cdot {\bf s}_k \; , 
\end{equation}
where the sum on $\langle i,j \rangle$ runs over all NN bonds and the sum on 
$\{i, k\}$ runs only over the selected NNN bonds shown in Fig.~\ref{fig1}.
In what follows we  
set $J_1 = 1$ and consider the case of (frustrating) $J_2 > 0$.
Although the GS of this model is well understood
in both the limits of small $J_2$ and large $J_2$, the GS phase at 
intermediate values of $J_2$ 
is still a matter of discussion.\cite{Mila,
Miyahara,Lauchli,Weihong,Weihong02,Muller,Kawakami,Hajii,Subir,Miyahara03,
Richter04,rachid05}
The CCM treatment of this model, briefly reviewed in this Section,  
is explained in more detail in Ref.~[\refcite{rachid05}].

We start with the classical 
GS  of the Shastry-Sutherland model. It 
is the collinear N\'eel state
for 
$J_2/J_1 \le 1$, but a 
non-collinear spiral state  for $J_2/J_1 > 1$ (and see Fig.~\ref{fig1} and
Refs.~[\refcite{Mila,Weihong02}]), 
with a characteristic 
pitch angle $\alpha_{\rm cl}$ given by
$\alpha_{\rm cl} =
\pi$ for $J_2 \le J_1$ and $\alpha_{\rm cl} = 
\pi - \arccos (J_1 / J_2)$ for   $J_2 > J_1$.
The transition from the collinear \Neel state to the 
non-collinear spiral state at $J_2/J_1=1$ is of second order for the classical model.
We note further 
that there are only two different angles between interacting spins, 
namely  
$\alpha_{\rm cl}$ for the $J_1$ couplings and $-2\alpha_{\rm cl}$ for the $J_2$ couplings.

\begin{figure}[ht]
\begin{center}
  \parbox{2.4in}{\epsfig{figure=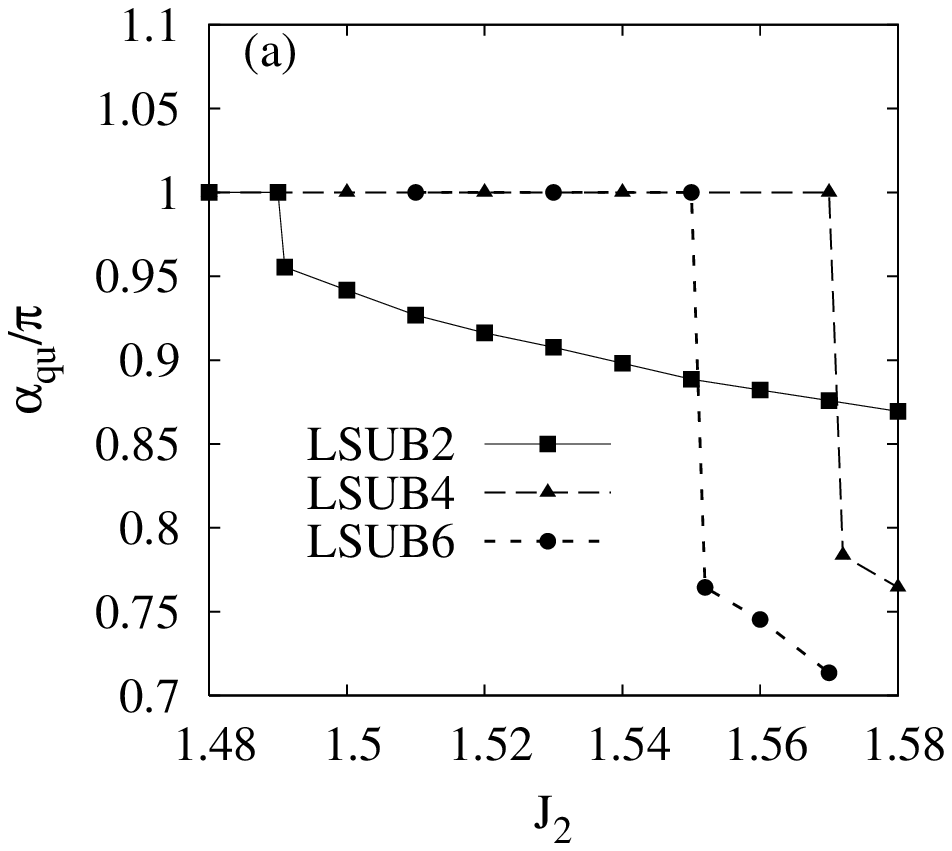,width=2.6in}}
  \hspace*{4pt}
  \parbox{2.4in}{\epsfig{figure=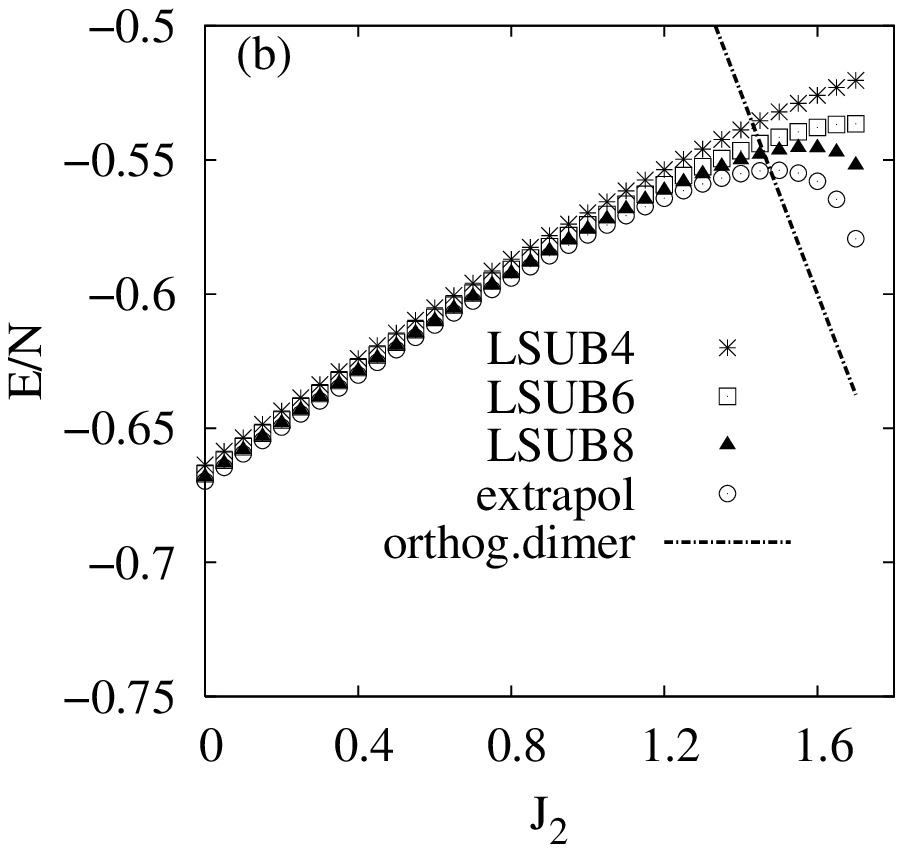,width=2.6in}}
\caption{(a) The quantum pitch angle $\alpha_{\rm qu}$ for the Shastry-Sutherland model 
with $J_1 = 1$ as a function of $J_2$, calculated within the
CCM-LSUB$n$ approximation with
$ n=2, 4, 6 $; and (b) the GS energy per spin as a function of $J_2$ of 
(i) the collinear \Neel phase  
obtained within the CCM LSUB$n$ approximation scheme with 
$n=4, 6, 8$ and its extrapolated $n \to \infty$ value using
Eq.~(\ref{scal_e}), and (ii) the orthogonal-dimer state.}
  \label{fig3}
\end{center}
\end{figure}

Similarly as we did for the quasi-1d $J_1$--$J_2$ model   
we now calculate the GS energy as a function of $J_2$
using as reference state a spiral state as shown in Fig.~\ref{fig1}.  
Again, we find that due to quantum fluctuations  the onset of the spiral phase in the quantum
model is shifted to higher values of $J_2$, and the transition 
between the collinear and 
the spiral states becomes discontinuous, as seen from Fig.~\ref{fig3}a.  
We conclude, that the collinear \Neel 
state is the favored CCM reference state up to values $J_2  \approx 1.5 J_1$.     

\def\figsubcap#1{\par\noindent\centering\footnotesize(#1)}

It is known from the early paper of Shastry and Sutherland\cite{Shastry}  
that for large  $J_2$ the
quantum GS of the model of Eq.~(\ref{eq1}) is a
rotationally-invariant 
orthogonal-dimer state given by
$|\Psi\rangle_{\rm dimer}  = 
\prod_{\{i,j\}_{J_2} } [|\uparrow_{i} \rangle|\downarrow_j \rangle
-|\downarrow_{i} \rangle|\uparrow_j \rangle]/\sqrt{2}\;$,
where 
$i$ and $j$ correspond to those sites which cover the $J_2$ bonds.
The energy per site of this state  
is $E_{\rm dimer}/N= -3J_2/8$. 
We compare its energy in Fig.~\ref{fig3}b with that of the CCM GS obtained with the collinear
\Neel state as reference state. Our results demonstrate that the 
orthogonal-dimer state has lower energy than the \Neel phase for
$J_2 \gtrsim 1.477J_1$, i.e., significantly before the point where the spiral CCM phase has lower energy than the collinear 
CCM phase. 
We note that we also checked that  $|\Psi\rangle_{\rm dimer}$ similarly 
remains the state of lowest energy in the region where the
non-collinear spiral phase has lower energy than the \Neel phase.
We conclude that there is no spiral GS phase 
in the quantum model. 
The critical value  $J_2^{d}=1.477J_1$ 
where the transition to the orthogonal-dimer 
phase takes place obtained by the CCM is in good agreement with results obtained by other methods (and 
see, e.g., Table 2 in Ref.~[\refcite{Miyahara03}]).

\begin{figure}[ht]
\begin{center}
  \parbox{2.4in}{\epsfig{figure=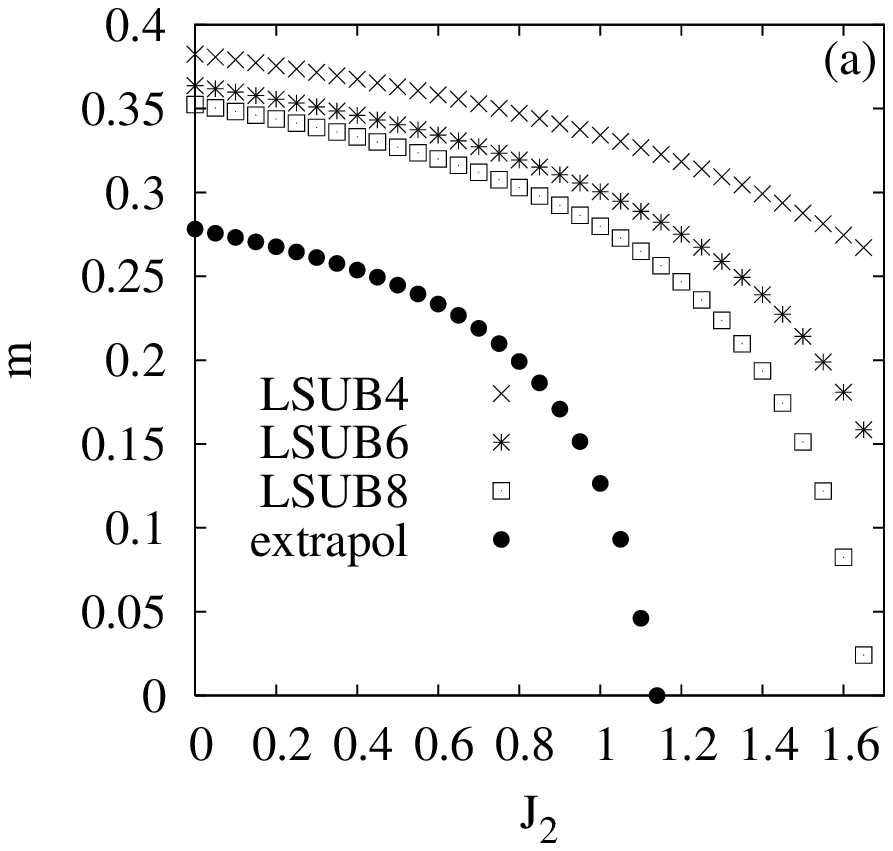,width=2.6in}}
  \hspace*{4pt}
  \parbox{2.4in}{\epsfig{figure=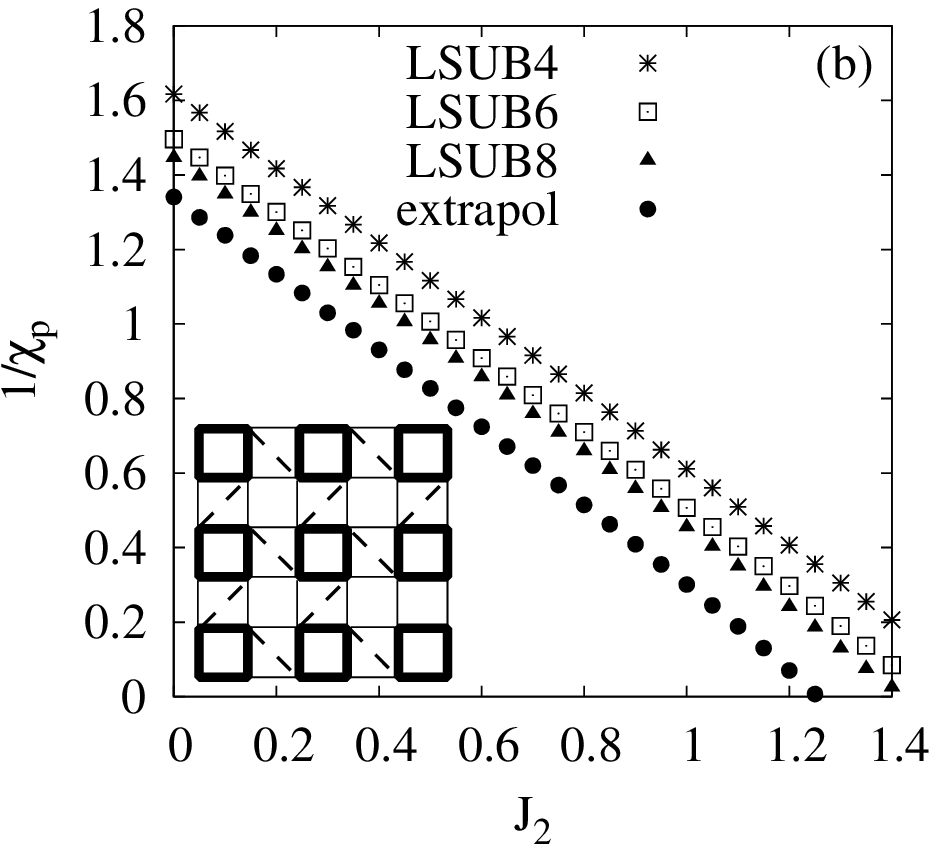,width=2.4in}}
\caption{(a) The sublattice magnetization $m$ versus $J_2$ 
obtained within the CCM LSUB$n$ approximation with 
$n=4, 6, 8$ and extrapolated to $n \to \infty$ 
using Eq.~(\ref{scal_m2}); and (b) the inverse susceptibility 
$1/\chi_p$ versus $J_2$ obtained within the CCM LSUB$n$ approximation with 
$n=4, 6, 8$  and extrapolated to $n \to \infty$ 
using Eq.~(\ref{scal_m1}). Inset: Pattern of plaquette valence bond order.}
  \label{fig5}
\end{center}
\end{figure}

So far we have mainly 
discussed the energy of competing GS phases.
The next question we would like to discuss is the question of the stability of
the \Neel LRO in the frustrated regime. For that purpose we calculate the order
parameter (viz., the sublattice magnetization) $m$
within the LSUB$n$ approximation scheme up to $n=8$ 
and extrapolate to 
$n \to \infty$ using the  extrapolation scheme of Eq.~(\ref{scal_m2}). 
The results are shown in 
Fig.~\ref{fig5}a.
The extrapolated data 
clearly demonstrate that the LRO vanishes before the 
orthogonal-dimer state becomes the GS. 
The transition from \Neel LRO to
magnetic disorder is of second order.
Hence, in agreement with previous investigations (and see, e.g., Ref. [\refcite{Miyahara03}]),  
we come to the second important conclusion that there exists 
an intermediate magnetically disordered phase.
The critical value  $J_2^{c}  \approx 1.14 J_1$ where the \Neel LRO breaks down
agrees well
with the corresponding value calculated by series expansion techniques as  
given, for example, in Table 2 of Ref.~[\refcite{Miyahara03}].

\def\figsubcap#1{\par\noindent\centering\footnotesize(#1)}

The nature  of the nonmagnetic GS in the region $1.14 J_2 \lesssim J_2 \lesssim 1.477J_2$
is a matter of some controversy in the literature.  A favored candidate is a  
plaquette singlet phase.\cite{Weihong02,Lauchli} To address this 
question we follow similar reasoning to that used in Ref.~[\refcite{sirker06}] and consider
the response of the spin system to
a field $F_p$ given by 
\begin{equation} \label{susc}
F_p = \delta \sum_{x, y} \big[(-1)^x{\bf s}_{x, y} \cdot {\bf s}_{x+1, y} + (-1)^y{\bf s}_{x, y} \cdot {\bf s}_{x, y+1}\big ] \; ,
\end{equation}
where $x,y$ are components (integer numbers) of the lattice vectors of the square lattice. 
This field corresponds to a  plaquette valence bond order (as illustrated in the inset of Fig.~\ref{fig5}b), which breaks 
the lattice symmetry.  Thus, 
we use the CCM with the \Neel state as reference state to  
calculate the energy per site $e(J_1,J_2,\delta)$ for $H+F_p$, namely the Hamiltonian 
of Eq.~(\ref{eq1}) perturbed by the additional term of Eq.~(\ref{susc}). 
The susceptibility $\chi_p$ is then defined as 
\cite{sirker06}  
\begin{equation}
\chi_p = -\left.\frac{\partial^2{e}}{\partial {\delta}^2} \right|_{\delta=0} \, .
\end{equation}
In Fig.~\ref{fig5}b we present the results for the inverse susceptibility, $1/ \chi_p$, as a function of $J_2$. 
The extrapolation to $n \to \infty $ 
is performed using an extrapolation scheme  with leading power $1/n$ as in Eq.~(\ref{scal_m1}).  
Clearly, in the magnetically ordered \Neel phase $\chi_p$ is finite as it should be. However, close to the transition to
the magnetically disordered phase at $J_2^{c}  \approx 1.14 J_1$ the susceptibility becomes very large and diverges at 
$J_2  \approx 1.26 J_1$,  which is close to 
$J_2^c$.
Hence, we conclude from our CCM data that there exists a valence-bond phase between the 
N\'{e}el-ordered 
phase and the orthogonal-dimer phase.

\section{The Quantum $J_1$--$J_2$ Antiferromagnet on the Stacked Square Lattice}
\label{3d}
As already mentioned in Sec.~\ref{intro} 
the $J_1$--$J_2$ model on the square lattice is a canonical model to study quantum phase transitions 
in $d=2$.\cite{chandra88,dagotto89,schulz,richter93,bishop98,
singh99,sushkov01,capriotti00,capriotti01,capriotti01a,siu01,singh03,sirker06}
However, in experimental realizations of the $J_1$--$J_2$ model
the magnetic couplings are expected to be not strictly 2d,
since a nonzero interlayer coupling $J_\perp$ is always present.
For example, recently Rosner {\it et al.}\cite{rosner02}
have found  $J_\perp \approx 0.07J_1$  for Li$_2$VOSiO$_4$, a 
material which can be described by a square lattice $J_1$--$J_2$ model 
with large $J_2$.\cite{melzi00,rosner02} 

Therefore, in this Section we consider the influence of an interlayer coupling on 
the GS  phases of the the $J_1$--$J_2$ spin-$1/2$ HAFM, i.e., we consider  
the  HAFM on the stacked square lattice described by
\begin{equation}
\label{ham}
H=\sum_n\Bigg(J_1\sum_{\langle ij \rangle}{\bf s}_{i,n} \cdot {\bf s}_{j,n}
+J_2 \sum_{[ ij ]}{\bf s}_{i,n} \cdot {\bf s}_{j,n}\Bigg)
 + J_\perp \sum_{i,n} {\bf s}_{i,n} \cdot {\bf s}_{i,n+1} \; ,
\end{equation}
where $n$ labels the layers 
and  $J_\perp \ge 0$ is the interlayer coupling.
The expression in parentheses 
represents the 
$J_1$--$J_2$ model of the layer $n$ with intralayer NN bonds 
$J_1=1$ 
and NNN bonds $J_2 \ge 0$.

The classical GS's of the model are the \Neel state for $J_2 < 0.5 J_1$ and another 
particular collinear state for 
$J_2 > 0.5 J_1$.  The latter state (which we henceforth refer to as the collinear-columnar or, simply the collinear state) 
is a columnar $(\pi ,0)$ state characterized by a parallel spin orientation of
nearest neighbors along the direction of one axis (say, the vertical or columnar direction) in each layer, and an antiparallel
spin orientation of nearest neighbors along the perpendicular (say, horizontal or row) direction. 
It is well known\cite{chandra88,dagotto89,schulz,richter93,bishop98,
singh99,sushkov01,capriotti00,capriotti01,capriotti01a,siu01,singh03,sirker06} that for $J_\perp=0$ 
the quantum model has two corresponding GS phases with semi-classical magnetic
LRO, one ({\Neel}-like) for small $J_2 \lesssim 0.4J_{1}$ and one (collinear-columnar-like) 
for large $J_2 \gtrsim 0.6 J_{1}$, which are
separated by a  magnetically disordered (quantum paramagnetic)
GS phase. 

\begin{figure}[ht]
\begin{center}
  \parbox{2.4in}{\epsfig{figure=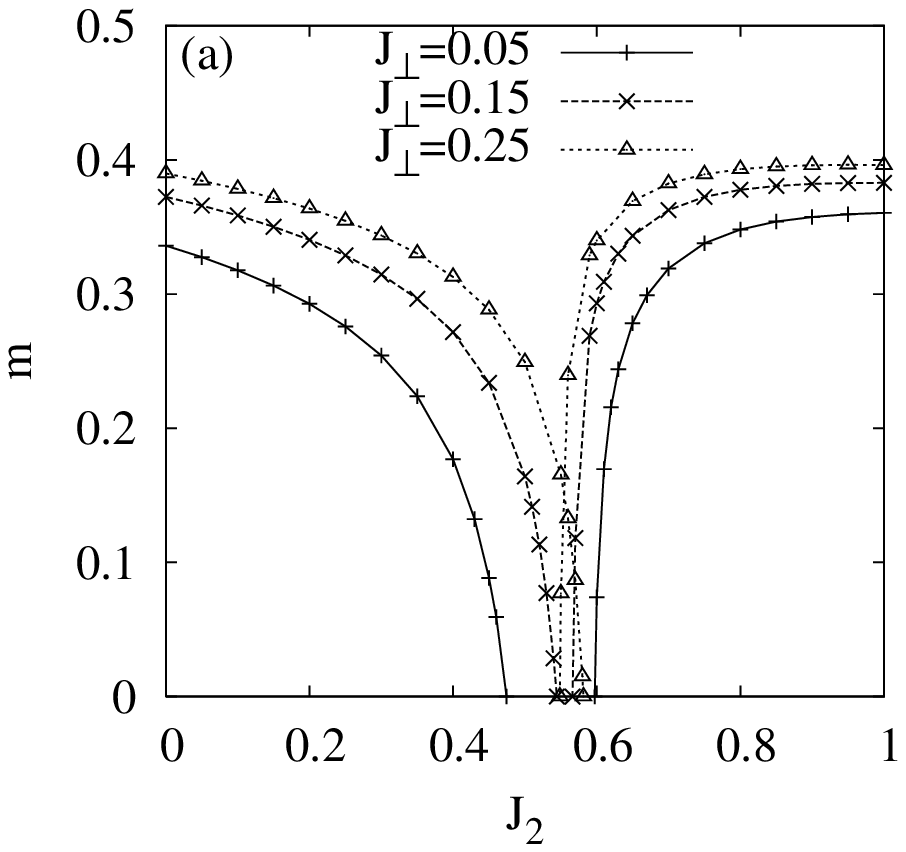,width=2.6in}}
  \hspace*{4pt}
  \parbox{2.4in}{\epsfig{figure=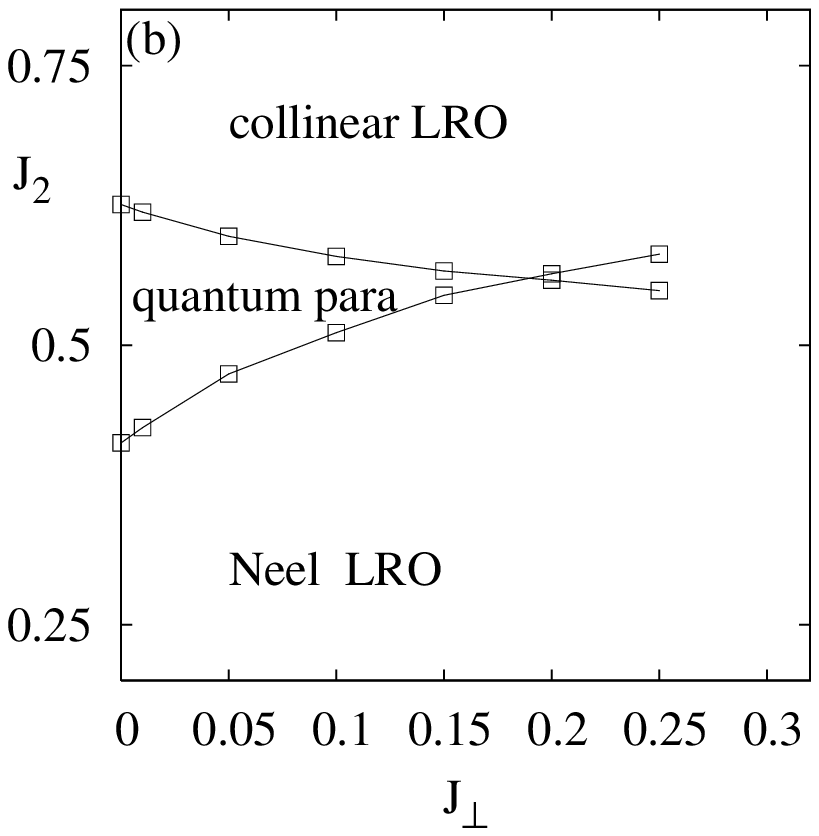,width=2.6in}}
\caption{(a) The magnetic order parameter $m$ versus $J_2$ for various strengths of
the interlayer coupling $J_\perp$ (with $J_1 = 1$); and 
(b) the ground-state phase diagram  
(where the solid lines show those values of $J_2$ for which the order
parameters vanish). Note that $J_1 = 1$.}
  \label{fig7}
\end{center}
\end{figure}

For the treatment of the model of Eq.~(\ref{ham}) with arbitrary $J_\perp$ we apply the CCM  
and 
use both classical GS's (\Neel and collinear-columnar) as reference states.
Here we illustrate the CCM approach to this model only very briefly, and refer the 
interested reader to Ref.~[\refcite{schmal06}] for more details.
In order to determine the GS phase transition points we calculate the order parameters 
for various values of $J_\perp$ 
and determine those values $J_2=\gamma_{\rm Neel}(J_\perp)$ and 
$J_2=\gamma_{\rm col}(J_\perp)$ where the order parameters vanish.
In Fig.~\ref{fig7}a we present some typical curves showing the extrapolated order parameters (according to Eq.~(\ref{scal_m2})) 
versus  $J_2$ for some values of $J_\perp$.  The magnetic order
parameters of both magnetically long-range ordered phases 
vanish continuously as is typical for second-order
transitions.  We note, however, that there are arguments\cite{schulz} 
that 
the transition from the collinear-columnar phase to the quantum paramagnetic phase
should be of first order.  
The order parameters for both phases 
are monotonically increasing functions of $J_\perp$, and the transition points 
$\gamma_{\rm Neel}$ and $\gamma_{\rm col}$ also move together as $J_\perp$ increases.
In Fig.~\ref{fig7}b we present the dependence on $J_\perp$ of these
transition points.  
Close to the strictly 2d case (i.e., for small $J_\perp \ll J_1$) the
influence of the interlayer coupling is largest.
For a characteristic value of $J^*_\perp \approx 0.19J_1$ 
the two transition points $\gamma_{\rm Neel}$ and $\gamma_{\rm col}$ 
meet each other. 
Hence, we conclude that already for quite weak interlayer coupling the magnetically disordered 
quantum phase that is present in the strictly 2d model disappears, and a direct transition between the two 
semi-classically ordered magnetic phases occurs.  This conclusion is also in agreement with the statement that 
there is no magnetically disordered phase in the 3d $J_1$--$J_2$ model on the bcc lattice.\cite{schmidt02,oitmaa04}   

\def\figsubcap#1{\par\noindent\centering\footnotesize(#1)}

\end{document}